  \documentclass[preprint,showpacs,preprintnumbers,amsmath,amssymb]{revtex4}

\usepackage{graphicx}
\usepackage{dcolumn}
\usepackage{bm}
\usepackage{subfigure}

\usepackage[utf-8]{inputenc}

\begin{document}

\title{Superdiffusion in quasi-two-dimensional Yukawa liquids}

\author{T. Ott}
\affiliation{%
    Christian-Albrechts-Universit\"at zu Kiel, Institut f\"ur Theoretische Physik und Astrophysik, Leibnizstra\ss{}e 15, 24098 Kiel, Germany
}%
\author{Z. Donk\'o}\author{P. Hartmann}
\affiliation{
Research Institute for Solid State Physics and Optics, Hungarian Academy of Sciences, P. O. Box 49, H-1525 Budapest, Hungary
}%
\author{M. Bonitz}%
\affiliation{%
    Christian-Albrechts-Universit\"at zu Kiel, Institut f\"ur Theoretische Physik und Astrophysik, Leibnizstra\ss{}e 15, 24098 Kiel, Germany
}%

\date{\today}

\begin{abstract}
The emergence and vanishing of superdiffusion in quasi-two-dimensional Yukawa systems are investigated by 
molecular dynamics simulations. Using both the asymptotic behaviour of the mean-squared displacement of the particles and the 
long-time tail of the velocity autocorrelation function as indicators for superdiffusion, we confirm 
the existence of a transition from normal diffusion to superdiffusion in systems changing from a three-dimensional 
to a two-dimensional character. A connection between superdiffusion and dimensionality is established by the behaviour of the projected pair 
distribution function. 
\end{abstract}

\maketitle

\section{Introduction}
Diffusion is a fundamental transport mechanism which plays a dominant 
role in many physical, chemical and biological systems. It is not 
only of academic but also practical interest to study diffusion in 
two-dimensional systems since many real-world systems can be described 
as being two-dimensional or quasi-two-dimensional, including surfaces or layers 
of small width, e.g. quantum wells. 
Two-dimensional diffusion has long been known to exhibit anomalous 
behaviour for a number of interactions and systems. One of the possible 
anomalies is the so-called superdiffusion which
describes diffusion proceeding faster than normal diffusion in 
the sense that a particle achieves a greater distance from its starting 
point than expected from Fick's law. 
For three-dimensional simple systems in thermodynamical equilibrium, 
to the best of our knowledge, 
no superdiffusion has been observed to 
date. On the other hand, under nonequilibrium conditions, anomalous 
transport is well-known, e.g. for chaotic systems \cite{zaslavsky2002cfk}, 
turbulent flows \cite{hauff2007ing}, 
or plasmas in turbulent magnetic fields \cite{pommois:012311}. These will not be considered here. 
For systems which are neither three-dimensional nor strictly 
two-dimensional, we expect a gradual transition from superdiffusive behaviour 
to Fickian diffusion. 

In two-dimensional systems of hard disks, a slow $\propto t^{-1}$ 
decay of the long-time tail of the velocity autocorrelation function (VACF) 
was first observed by Alder and Wainwright \cite{alder1970dva}. Such a decay results in a 
divergent Green-Kubo integral. Superdiffusion was found to take place 
in quasi-two-dimensional complex plasmas \cite{ratynskaia:105010, QuinnGoree2002, LaiI2002, JuanI1998}, 
including driven-dissipative systems \cite{liu:055003} and systems under laser-induced shear \cite{JuanChenI2001}.
No experimental evidence for superdiffusion was found by Nunomura et al. who examined 
an underdamped liquid complex plasma \cite{nunomura:015003}. 

The intent of this study is to examine the superdiffusion in systems 
which are quasi-two-dimensional, i.e. the extension in one spatial dimension of the system 
is much smaller than in the other two. This is often the situation in 
experimental setups, for example in dusty plasma experiments where the 
dust grains are levitated by an electrostatic force which is counteracted 
by gravity. It is clear that under such circumstances 
the particles are not rigorously restricted to a two-dimensional plane 
but form a quasi-two-dimensional system. For example, Ref. \cite{QuinnGoree2002} reports 
on experiments in which the dusty plasma under consideration consisted of two to three layers. 

Here we consider a macroscopic system of charged particles interacting via a
screened Coulomb (Yukawa) potential. It is of relevance to 
dusty plasmas \cite{bonitz2006sps}, colloidal suspensions, electrolytes and other systems.
It is also an important theoretical tool since it allows to tune the inter-particle interaction 
(by varying the screening length of the Yukawa potential) 
from being very long-ranged to almost contact interaction. 

We report on molecular dynamics studies performed for quasi-two-dimensional 
systems. Our 
focus lies on the influence of the degree of quasi-two-dimensionality 
on the vanishing and emergence of superdiffusion.

\section{Model and Simulation technique}
We study the system using equilibrium molecular dynamics 
simulations (e.g. \cite{golubnychiy2001dpa}). The interaction of
the particles is 
given by the Yukawa pair potential
\begin{equation}
\phi(r)=\frac{Q}{4\pi\varepsilon_0}\frac{e^{-r/\lambda_D}}{r}
\label{eq:yukawa}
\end{equation}
Here, $Q$ is the particles' charge and $\lambda_D$ is the screening 
length. \\
A two-dimensional Yukawa system is characterized by two parameters, 
the Coulomb coupling parameter $\Gamma=(Q^2/4\pi\varepsilon_0)\times(1/a_{ws}k_BT)$
and the screening parameter $\kappa = a_{ws}/\lambda_D$. 
$T$ is the temperature and $a_{ws}=(n\pi)^{-1/2}$ is the Wigner-Seitz 
radius for two-dimensional systems. 
While our simulation box is periodic in the $x$ and $y$ directions, it is unbound in 
the $z$ direction. The particles' perpendicular movement in 
$z$ direction is restricted by one of two confinement potentials
\begin{eqnarray}
V^{\textrm{harm}}(z)&=&f\frac{Q^2}{4\pi\varepsilon_0 a_{ws}^3} \frac{z^2}{2}
\label{eq:harmonic}\\
V^{\text{box}}(z)&=&\frac{Q^2}{4\pi\varepsilon_0}\left ( \frac{e^{-\left ( z+W\right )/\lambda_D}}{z+W}  + \frac{e^{-\left ( -z+W\right )/\lambda_D}}{-z+W} \right ) 
\label{eq:soft-box}
\end{eqnarray}

The confinement (\ref{eq:harmonic}) is a simple harmonic trap where 
$f$ denotes the trap amplitude \cite{donko2004cmq}. 

\begin{figure}
\scalebox{0.68}{\includegraphics{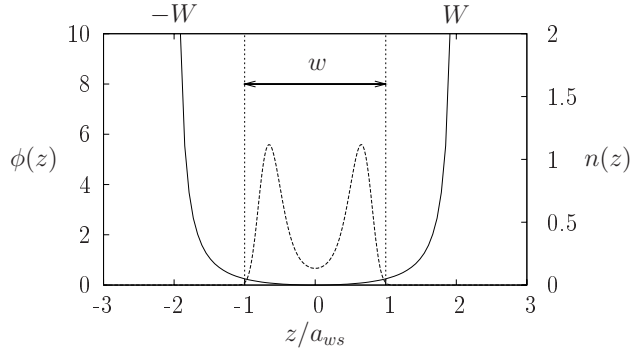}}
\caption{\label{fig:trap_illust} 
Soft-box potential $\phi(z)$ and corresponding particle density $n(z)$. 
The trap is impenetrable at $\pm W$ and the particles can typically explore the trap width $w$ (here, $w=2a_{ws}$). }
\end{figure}

The second model, Eq. (\ref{eq:soft-box}), uses a box-shaped confinement with ``soft'' walls. 
It can be thought of as consisting of two rigid walls of %indisplaceable 
immovable
particles with the same interaction as the actual particles, i.e.
a Yukawa interaction, see Fig. \ref{fig:trap_illust} for an illustration. Here, $W$ is defined
as $W=  \frac{w}{2}+a_{ws}$ with $w$ being the ``width'' 
of the box, i.e. twice the maximum displacement in $+z$ or $-z$ 
direction. We use the width $w$ as a third parameter for the soft-box 
confinement. \\
The choice between two confinements allows us to (a) reproduce typical 
experimental situations or (b) study the effect of the dimensionality on 
superdiffusion while retaining comparable plasma conditions at all times. 
We achieve this by the following procedure: In the harmonic confinement, 
as the amplitude of the trap is decreased, the particle number is left 
unchanged and the particles can explore a wider vertical space. This 
results in an increased mean inter-particle distance, i.e. the 
(three-dimensional) density is decreased. 
Contrarily, in the soft-box confinement, we chose $w=a_{ws}$ as a 
point to fix for the 3D density and scale the particle number 
as necessary to maintain that density for all values of $w$. \\

We performed simulations for a fixed coupling parameter $\Gamma=200$. 
Prior to measurement, the particles' velocities are rescaled at 
each timestep to the desired temperature until a Maxwell distribution 
is well-established. During the measurement, the velocities are not 
rescaled. 
Our simulations are carried out for $\kappa = 2.0$ and 
$\kappa = 3.0$. At these parameters, a 2D Yukawa system is 
well in the liquid phase, with the melting points being 
$\Gamma \approx 415$ and $\Gamma \approx 1210$ for 
$\kappa = 2.0$ and $3.0$ respectively \cite{hartmann2005epa}.\\
The particle number in our simulations is $N=6000$ for the 
harmonic confinement and $N=4000\dots16000$ for the soft-box 
confinement with $N\geq6000$ for $w\geq1.5 a_{ws}$. \\
In the following, time is given in units of the inverse plasma frequency 
$\omega_p = (Q^2/2\pi\varepsilon_0ma_{ws}^3)^{1/2}$ with $m$ being the 
mass of the particles and lenghts are measured in units of $a_{ws}$. 
We solve the equation of motion for each particle using 
the velocity Verlet algorithm \cite{allen1987csl}. 

% \section{Superdiffusion}
The structure of the systems is characterized by measuring the density distribution  $n(z)$
perpendicular to the confined direction and by the 
\emph{projected pair distribution function} $g^\ast(r)$. \\
To analyze diffusion properties we calculate the time-dependence of the mean-squared 
displacement (MSD)
\begin{equation}
 u_r(t)=\langle \vert \vec r(t) - \vec r(t_0) \vert^2 \rangle
\end{equation}
where $\langle . \rangle$ denotes an ensemble-average and 
$\vec{r} = (x \; y )$ is the position vector in the plane; the $z$-component 
is bound from above due to the confining force and does not 
need to be taken into account. \\
The motion can be classified according to the time-dependence 
$u_r(t)\sim t^\gamma$. Normal Fickian diffusion is characterized 
by a linear time dependence, $\gamma=1$. If $\gamma > 1$ or
$\gamma < 1$, motion is super- or subdiffusive, respectively. 
Ballistic, i.e. undisturbed, motion is trivially marked by 
$\gamma = 2$. \\
We have carried out calculations of the MSD for different 
trap amplitudes and box widths and determined the slope of
the curve on a double logarithmic plot between 
$t=100 \omega_p^{-1}$ and $t=300 \omega_p^{-1}$ which yields
the diffusion exponent $\gamma$.

A more direct insight into a particle's movement can be obtained 
from the decay of the velocity autocorrelation function (VACF)
\begin{equation}
 Z(t) = \langle \vec v(t) \cdot \vec v(t_0)  \rangle
\end{equation}
where $\vec v = (v_x \; v_y )$. 
$Z(t)$ is a measure of the memory of the system. For uncorrelated 
binary collisions, $Z(t)$ is expected to decay exponentially. 
If $Z(t)$ decays algebraically, $Z(t)\sim t^{-\alpha}$, $\alpha$ 
needs to be larger than 1 for diffusion to be Fickian, because 
otherwise no valid diffusion coefficient can be calculated from the 
Green-Kubo formula. For 3D, the decay has been found to be 
algebraic with $\alpha=1.5$ for a number of pair potentials in 
experiments 
\cite{paul1981olt,PhysRevLett.58.1873,PhysRevLett.68.2559}, 
simulations  
\cite{alder1970dva,PhysRevE.63.026109}
and by theoretical models \cite{pomeau1975tdc,PhysRevA.2.2005}. 

% cite Alder und Wainwright? \cite{alder1970dva}
For 2D Yukawa systems exhibiting superdiffusion, Liu and Goree 
have found an algebraic decay with $\alpha \approx 1$ \cite{liu2007std}. 
This is an indication of superdiffusive behaviour. 

In our simulations, we calculated the velocity autocorrelation 
for different trap amplitudes. To obtain accurate statistics, 
we performed between 10 and 50 runs of 700.000 timesteps for 
each trap amplitude and again determined the slope of the curve 
in a double logarithmic plot, this time between 
$t=100 \omega_p^{-1}$ and $t=250 \omega_p^{-1}$. 
Special attention must be paid to the error estimation. We used 
the Jackknife method to evaluate our data, because standard 
error estimates may not be sufficient in this case \cite{jackknife, shao1995jab}.

\section{Results}
\subsection{Harmonic confinement}
\begin{figure}
\scalebox{0.68}{\includegraphics{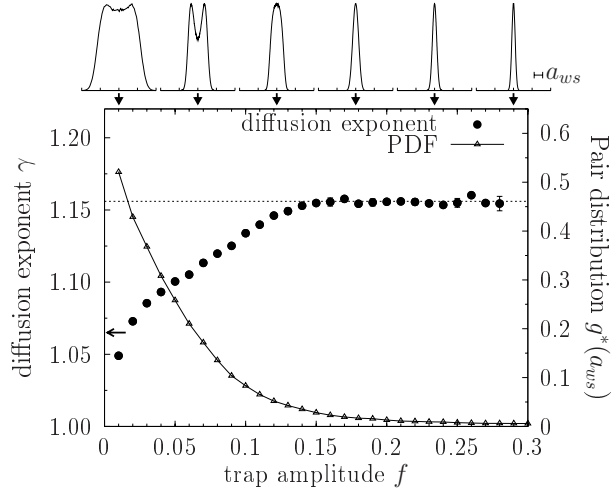}}
\caption{\label{fig:k2msd} The diffusion exponent for different
trap amplitudes and $\kappa = 2.0$ in the harmonic confinement.
The straight line is a guide for the eyes. 
Also shown is the value of the (projected) pair correlation function 
at the distance $r=a_{ws}$. The top graphs show the density 
profile in the confined direction from $z=-4a_{ws}$ to $z=4a_{ws}$ at the trap amplitude indicated by the arrows. 
The $n(z)$ distributions are normalized here to unit amplitude.  }
\end{figure}

\begin{figure}
\scalebox{0.68}{\includegraphics{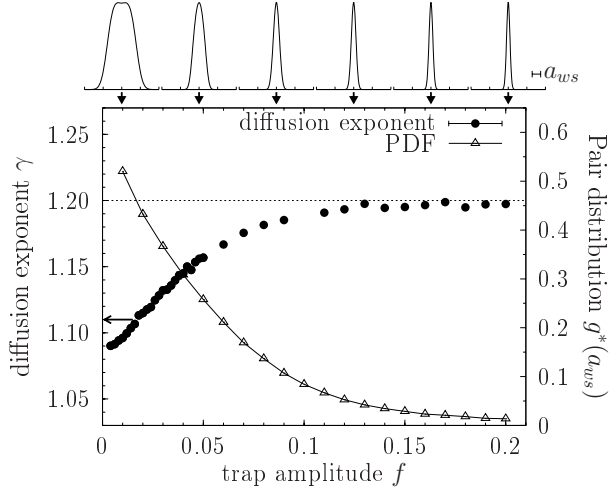}}
\caption{\label{fig:k3msd} The diffusion exponent for different
trap amplitudes and $\kappa = 3.0$ in the harmonic confinement. 
Also shown is the value of
the (projected) pair correlation function at the distance $r=a_{ws}$. 
The top graphs show the density profile from $z=-4a_{ws}$ to $z=4a_{ws}$ 
in the confined direction  at the trap amplitude indicated by the arrows. The $n(z)$ distributions are normalized here to unit amplitude. }
\end{figure}

We begin by noting that
particles interacting via a Yukawa potential and confined by a 
harmonic trap support the formation of layers \cite{donko2004cmq, PhysRevLett.78.3113}.
The number of layers formed depends on the temperature, the 
trap frequency and the screening length. For our choice of 
parameters, we found that for $\kappa=2.0$ a second 
layer builds up at $f=0.12$ while for $\kappa=3.0$ the 
system consists of a single layer for the whole range of $f$ 
examined (cf. the insets above Figs. \ref{fig:k2msd} and \ref{fig:k3msd}).

\begin{figure}
\scalebox{0.68}{\includegraphics{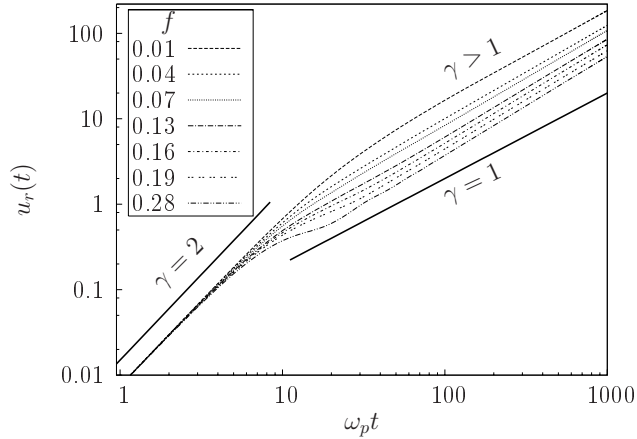}}
\caption{\label{fig:msd} The MSD over time for 
different trap amplitudes in the harmonic confinement and $\kappa = 2.0$.
The solid lines have slope 2.0 (ballistic regime) and 1.0 (normal diffusion)
respectively. The slope of the MSD in this log-log plot is the diffusion
exponent, see Fig. \ref{fig:k2msd}. }
\end{figure}

Typical results for the time-dependence of $u_r(t)$ are shown in 
Fig. \ref{fig:msd}. In this double-logarithmic plot, the slope of the 
curves corresponds to the exponent of the algebraic behaviour. 
For $t<10\omega_p^{-1}$, $u_r(t)$ grows quadratically with time, which 
corresponds to a ballistic motion of each individual particle. After a 
narrow transition region of about $10\omega_p^{-1}$, the particles' 
movement is dominated by diffusive processes. The slope in this 
region allows us to classify motion as superdiffusive, diffusive 
or subdiffusive, i.e. it is the diffusion exponent $\gamma$. 
In Fig. \ref{fig:msd}, $u_r(t)$ is depicted for different 
trap amplitudes $f$. For strong confinements, $u_r(t)$ deviates strongly
from a purely diffusive behaviour of $\gamma=1$. For increasingly more relaxed 
confinements, the slope of $u_r(t)$ tends more and more towards unity, that is
the migration becomes less superdiffusive.

The dependence of the diffusion exponent $\gamma$ on the trap amplitude $f$
is shown in Figs. \ref{fig:k2msd} and \ref{fig:k3msd}.

It is clear that the degree 
of superdiffusivity, as measured by the diffusion exponent $\gamma$, 
decreases with an increasing width of the system.
For $\kappa = 2.0$, Fig. \ref{fig:k2msd}, the diffusion exponent is in the vicinity 
of $\gamma=1.16$ for strong confinements which coincides with the value we obtained for 
strictly 2D systems. When the width of the particle distribution along 
the z-axis exceeds $2 a_{ws}$ (which is near the mean interparticle separation in the strictly		 2D case), 
the diffusion exponent begins to deviate from that value 
and continues to fall for lower trap amplitudes $f$. The same behaviour can 
be seen in Fig. \ref{fig:k3msd} for $\kappa = 3.0$. Here, $\gamma$ 
saturates at $\approx 1.20$ for strongly confined systems and again falls when 
the system width exceeds $2 a_{ws}$. 
As noted before, the 3D density of the system differs for different 
trap amplitudes. To exclude the possibility that the vanishing of superdiffusion 
is only an effect of the density, we also simulated strictly 2D systems in 
which we changed the particle density until the first peak in the pair 
distribution function coincided with that of our quasi-2D systems. Our data (not shown) 
clearly indicates that a decreased density does not lower the diffusion exponent 
in our parameter range. In fact, for $\kappa=2.0$ the superdiffusion is stronger 
for systems of lower density. 

To support the idea that the vanishing of superdiffusion is connected 
with the fact that particles can pass each other in z-direction, 
we analyze the projected pair distribution function $g^\ast(r)$. Its value 
at $r=a_{ws}$ is also shown in Figs. \ref{fig:k2msd} and \ref{fig:k3msd}. 
For nearly 2D systems (high $f$), $g^\ast(a_{ws})$ is practically zero. 
The reason for this is that for strongly correlated liquids, the particles' 
kinetic energy is not sufficient for two particles to come together as close as 
one Wigner-Seitz radius. Instead, they are trapped in local potential minima from which 
they can escape only after some time. 
Figs. \ref{fig:k2msd} and \ref{fig:k3msd} show that $g^\ast(a_{ws})$ 
is non-zero for lower trap amplitudes $f\lesssim 0.1$. This is a result of the 
projection of all particles onto a two-dimensional plane. Individual particles
are still separated by more than $a_{ws}$ except now the finite width 
of the system allows the particles to go ``over and under'' each other in $z$-direction 
and the particles' projections can be close. 

By inspection of Figs. \ref{fig:k2msd} and \ref{fig:k3msd}, 
we see that the behaviour of the dynamic quantity $\gamma(f)$ is 
closely mirrored by the static quantity $g^\ast(a_{ws})\vert_f$. 
This underlines that superdiffusion is connected with the dimensionality
of the system.

\begin{figure}
\subfigure{
\scalebox{0.68}{\includegraphics{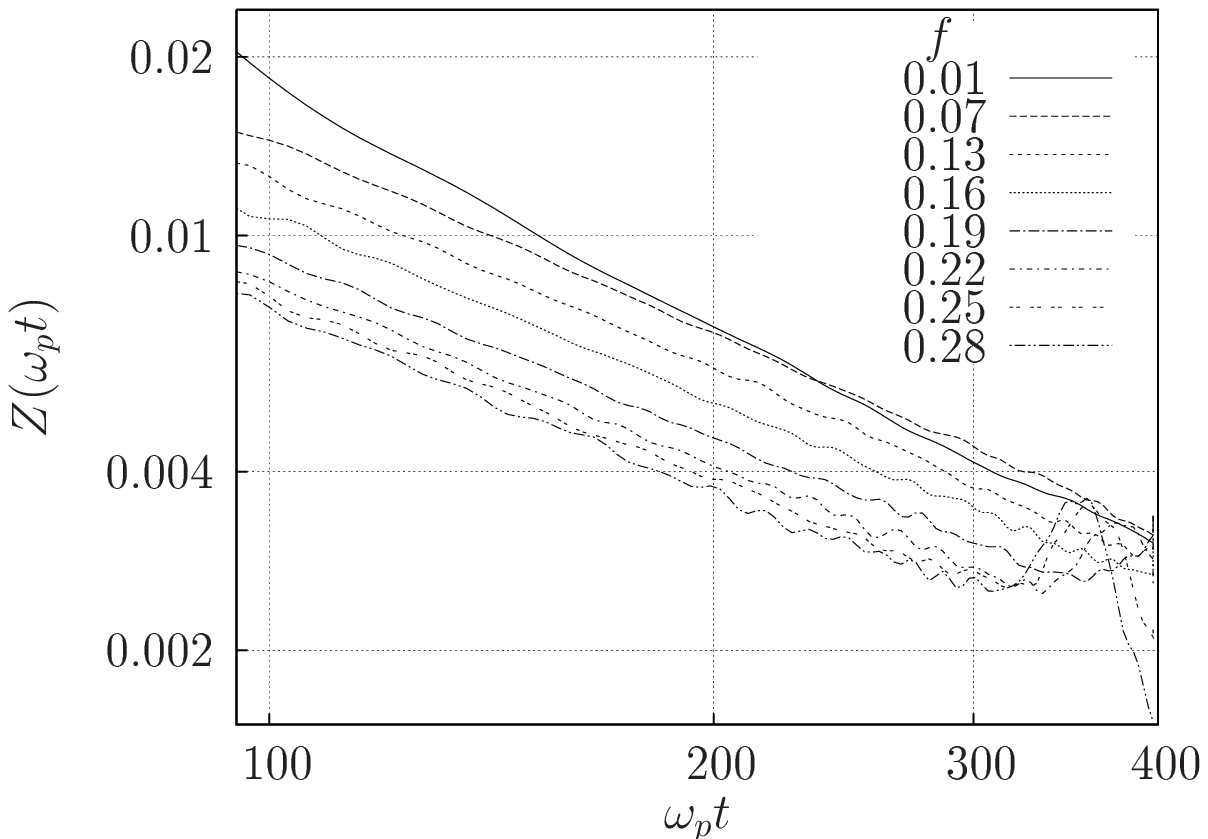}}
\label{fig:vacf_illust}
  }
 \subfigure{
\scalebox{0.68}{\includegraphics{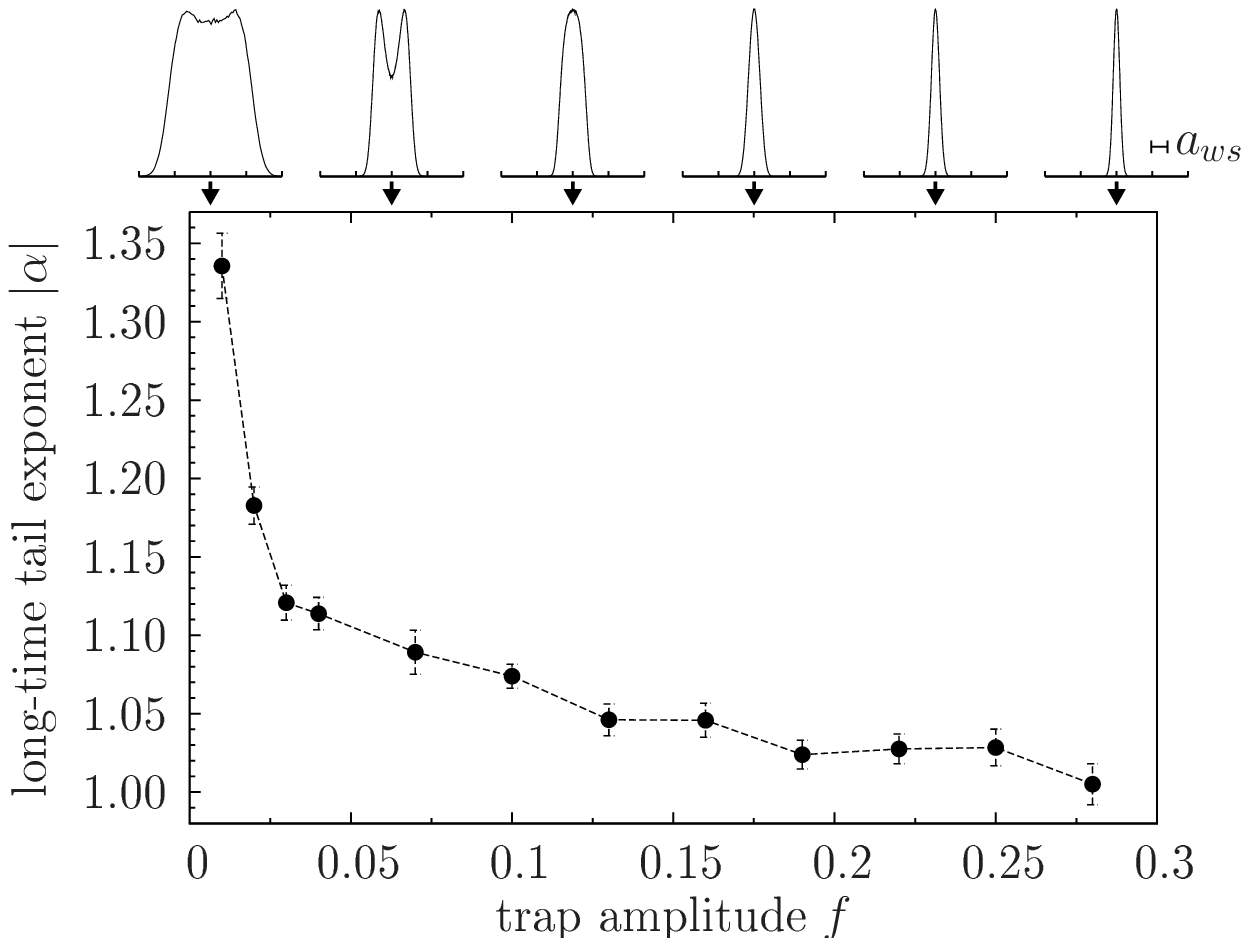}}
\label{fig:q2dvacf}
  }

\caption{ 
 \subref{fig:vacf_illust} Double-logarithmic plot of the long-time tail of the VACF $Z(t)$ ($Z(0)=1$) from $f=0.01$ (top)
to $f=0.28$ (bottom). \\
\subref{fig:q2dvacf} The exponent of the algrebraic decay of 
the velocity auto-correlation function for $\kappa = 2.0$
and different trap amplitudes  in the harmonic confinement. 
Error bars denote standard errors from the Jackknife estimator.  
The top graphs show the density profile in the confined direction 
from $z=-4a_{ws}$ to $z=4a_{ws}$ at the trap amplitude indicated by the arrows. 
The $n(z)$ distributions are normalized here to unit amplitude. } 
\end{figure}
\begin{figure}
\scalebox{0.68}{\includegraphics{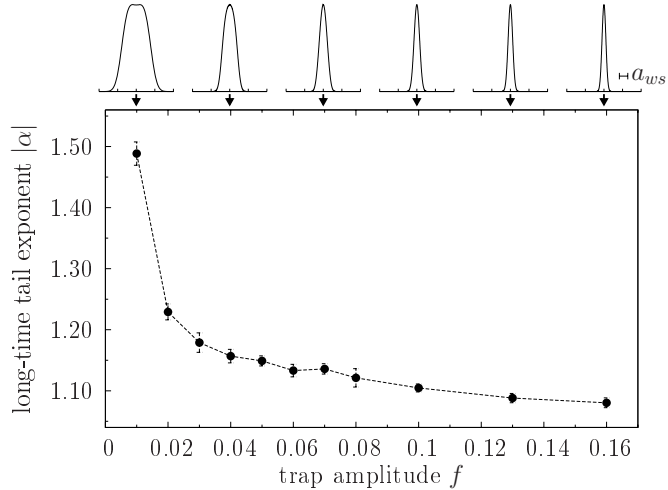}}
\caption{\label{fig:k3vacf} The exponent of the algrebraic decay of 
the velocity auto-correlation function for $\kappa = 3.0$
and different trap amplitudes  in the harmonic confinement. 
Error bars denote standard errors from the Jackknife estimator.  
The top graphs show the density profile in the confined direction 
from $z=-4a_{ws}$ to $z=4a_{ws}$ at the trap amplitude indicated by the arrows. 
The $n(z)$ distributions are normalized here to unit amplitude. }
\end{figure}

Fig. \ref{fig:vacf_illust} depicts the asymptotic behaviour of $Z(t)$ for different
trap amplitudes $f$ in a double logarithmic plot. 
The results for the decay of the VACF are shown in Figs. \ref{fig:q2dvacf} and \ref{fig:k3vacf} for $\kappa=2.0$ and
$\kappa=3.0$.

The data shown in Fig. \ref{fig:vacf_illust} validate our approach of modelling the asymptotic behaviour of $Z(t)$ 
as an algebraic decay since $Z(t)$ closely follows a straight line in the log-log plot. 
The peak seen in Fig. \ref{fig:vacf_illust} at long times for high $f$ is due to our finite simulation box and is caused 
by sound waves travelling through and re-entering the system due to the periodic boundary 
conditions. Measurement is limited to times smaller than the time of the sound wave 
traversal. 
The dependence of $\alpha$ on the trap amplitude $f$, Fig. \ref{fig:q2dvacf} and \ref{fig:k3vacf}, indicates 
vanishing of superdiffusion for broader systems. Starting close to 
the value $\alpha = 1.0$ for narrow systems, we see an 
increase in $\alpha$ for broader systems. This corresponds to a faster decay which 
is indicative of a loss of the particles memory, i.e. at subsequent times, a particle 
is less likely to travel in its original direction.

\subsection{Soft-box confinement}
We now turn our attention to the case of the soft-box confinement. 
Again, let us note that here too, the system forms layers when 
given enough space in the confined direction. We find that 
the number of layers formed is higher than in the harmonic 
confinement which we attribute to the constancy of the particle 
density. Recall that in this case we change the number of particles to 
maintain a constant 3D density. 

\begin{figure}
\scalebox{0.68}{\includegraphics{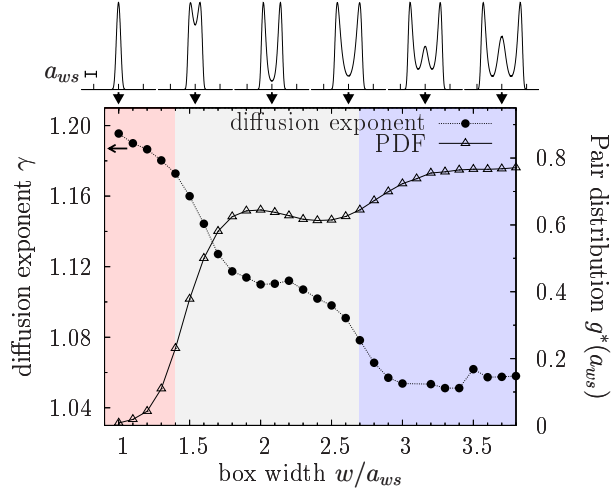}}
\caption{\label{fig:k3boxmsd} (color online) The diffusion exponent for different
box widths and $\kappa = 3.0$ in the soft-box confinement.
Also shown is the value of the (projected) pair correlation function 
at the distance $r=a_{ws}$. The top graphs show the density 
profile in the confined direction from $z=-3a_{ws}$ to $z=3a_{ws}$ at the box width indicated by the arrows. 
The $n(z)$ distributions are normalized here to unit amplitude. 
The background color separates regions of one, two and three layers. }
\end{figure}

\begin{figure}
\scalebox{0.68}{\includegraphics{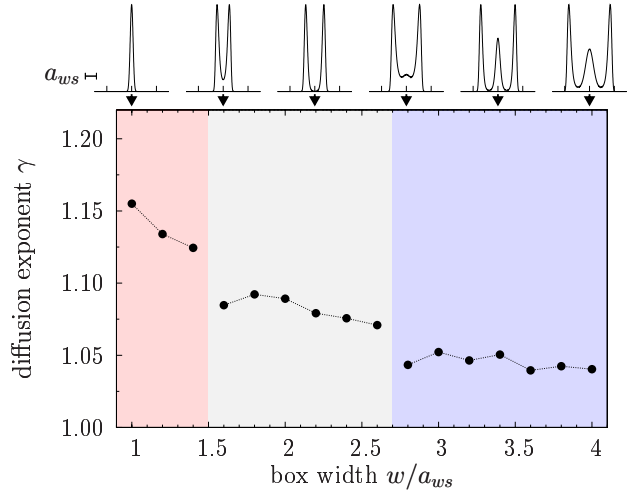}}
\caption{\label{fig:k2boxmsd} (color online) The diffusion exponent for different
box widths and $\kappa = 2.0$ in the soft-box confinement.
The top graphs show the density 
profile in the confined direction from $z=-3a_{ws}$ to $z=3a_{ws}$ at the box width indicated by the arrows. 
The $n(z)$ distributions are normalized here to unit amplitude. 
The background color separates regions of one, two and three layers. }
\end{figure}

The dependence of the diffusion on the width of the system, 
Figs. \ref{fig:k3boxmsd} and \ref{fig:k2boxmsd}, is more involved 
than for the harmonic confinement. Again, the general trend is 
for superdiffusion to vanish for increasingly broader systems. But 
here, the vanishing happens in stages: After a first drop, the 
diffusion exponent reaches a plateau from which it drops to a 
second plateau. This behaviour is connected to the formation of
layers in the system as indicated by the different background 
colors in Figs. \ref{fig:k3boxmsd} and \ref{fig:k2boxmsd} 
(cf. top graphs in these Figs). As another layer is formed 
in the system, the diffusion exponent experiences a drop. 
The non-monotonicity of the curve in Figs. \ref{fig:k3boxmsd} and \ref{fig:k2boxmsd} 
is due to statistical error.


We see that two to three 
layers are already sufficient to reduce the superdiffusive behaviour 
substantially. In addition, we again notice that the value of 
the projected pair distribution function $g^\ast(a_{ws})$ 
is directly correlated with the diffusion exponent.

\section{Summary and Discussion}
A study of superdiffusion in quasi-two-dimensional 
Yukawa liquids was performed by equilibrium molecular 
dynamics simulations. 
The two indicators for superdiffusion employed, the MSD 
and the VACF, both show sensivity to the dimensionality 
of the system. For increasingly broader systems
superdiffusion gradually vanishes. The transition from superdiffusion to 
normal diffusion was tested for two representative values of $\kappa$ and
found to be qualitatively comparable. This leads us to the conclusion 
that the transition is universal for Yukawa systems in the fluid phase. 

To ensure that the choice of confinement does not interfere with 
the change in dimensionality, we have used two different schemes 
to confine the system. The general trend of the vanishing of 
superdiffusion appears to be independent of the type of confinement. 
The finer details of how
the vanishing happens depend on the choice of confinement and here 
especially on the formation of layers. 

The strength of superdiffusion at zero width and the number of layers 
in the systems depend on the inter-particle potential and the system 
parameters. At a fixed Coulomb coupling parameter $\Gamma=200$, superdiffusion is stronger for 
$\kappa = 3.0$ and weaker for $\kappa =2.0$. 

By inspection of the projected pair distribution function we have established 
a close connection between superdiffusion and the dimensionality of the system. 
To this end, 
\begin{figure}
\scalebox{0.68}{\includegraphics{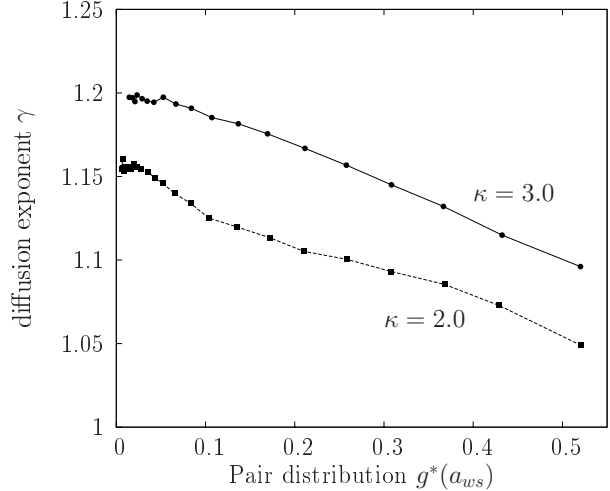}}
\caption{\label{fig:diff_vs_gofr} The diffusion exponent $\gamma$
over the value of the projected pair distribution $g^\ast(a_{ws})$ 
for $\kappa=2.0$ and $\kappa=3.0$ in the harmonic confinement. }
\end{figure}
we display in Fig. \ref{fig:diff_vs_gofr} the dependence of the diffusion exponent $\gamma$ on 
the value of the projected pair distribution $g^\ast(a_{ws})$ 
for $\kappa=2.0$ and $\kappa=3.0$ in the harmonic confinement, which is close to linear. For the 
soft-box confinement the general behaviour is similiar, although details are more complex due to the 
occurence of plateaus in the curves, cf. Fig. \ref{fig:k3boxmsd}. 

It remains an interesting question what types of other pair potentials also
support superdiffusion and how its strength depends on the interaction range.
Finally, it will also be of high interest for future analysis to see how
quantum effects influence superdiffusion. This could be done, e.g., by use of
effective quantum pair potentials \cite{filinov2003ikp,filinov2004tdq}.  

\begin{acknowledgements}
We acknowledge stimulating discussions with J.W. Dufty. This work has been supported by the Deutsche Forschungsgemeinschaft via SFB-TR24 grant A5 and by the Hungarian Scientific Research Fund, through grants OTKA-T-48389, OTKA-IN-69892 and PD-049991. 
\end{acknowledgements}

%\newpage 
%\bibliography{SD}

\begin{thebibliography}{28}
\bibitem[{jac()}]{jackknife}
\bibinfo{note}{Given $M$ samples $\{y_0(t), y_1(t), \dots y_M(t)\}$ of the
  VACF, we perform $M$ least-square fits of the average of $M-1$ of these
  samples to the target function $\alpha t+c$, leaving one different sample out
  from the average each time. The mean value $\bar{\alpha}$ is computed from
  the $M$ fit values and the error is estimated as the square root of
  \mbox{$\sigma^2 =(M-1) \sum_{i=1}^M (\bar{\alpha} - \alpha)^2/M$} where
  $\alpha$ is the result obtained by fitting over the average of all $M$
  samples.}
\end{thebibliography}

\end{document}